\documentclass[aps,prb,twocolumn,eqsecnum]{revtex4} 

\usepackage{graphicx}

\usepackage{epsf}

\newcommand{\re}[1]{(\ref{#1})}

\newcommand{\up}{\uparrow}

\newcommand{\dn}{\downarrow}

\newcommand {\dis}{\displaystyle}

\newcommand{\B}{B}

\newcommand{\g}{\textrm{g}}

\newcommand{\beg}{\begin{equation}}
\newcommand{\en}{\end{equation}}

\newcommand{\eps}{\epsilon}
\newcommand{\lam}{\lambda}

\begin{document}

\title{Solution for the dynamics of the BCS and central spin problems}  

\author{Emil A. Yuzbashyan$^{1}$ }

\author{Boris L. Altshuler$^{1, 2}$}

\author{Vadim B. Kuznetsov$^3$}

\author{Victor Z. Enolskii$^4$}

\affiliation{ \phantom{a}
\vspace{0.1cm}
\centerline{$^1$Physics Department, Princeton University, Princeton, NJ 08544, USA} 
\centerline{$^2$NEC Research Institute, Princeton, NJ 08540, USA}
\centerline{$^3$Department of Applied Mathematics, University of Leeds, Leeds, LS2 9JT, UK}
\centerline{$^4$Department of Mathematics, Heriot-Watt University, Edinburgh EH14 4AS, UK} }



 
\begin{abstract}
We develop  an explicit description of a time-dependent response of fermionic condensates to perturbations. The dynamics of Cooper pairs at times shorter than the energy relaxation time can be described by the BCS model. We obtain a general explicit  solution for the dynamics of the BCS model.  We also solve a closely related dynamical problem -- the central spin model, which describes a localized spin coupled to a ``spin bath''. Here, we focus on presenting the solution and describing its general properties, but also  mention some applications, e.g. to nonstationary  pairing in cold Fermi gases and  to the issue of electron spin decoherence in quantum dots. A typical dynamics of the BCS and central spin models is quasi-periodic with a large number of frequencies and stable under small perturbations. We show that for certain special initial conditions  the number of frequencies decreases and the solution simplifies. In particular, periodic solutions correspond to the ground state and excitations of the BCS model.

\end{abstract}

\maketitle    

\section{Introduction}

Recent experiments \cite{feshbach_experiments} on cold fermion pairing in the vicinity of a Feshbach resonance offer a unique opportunity to control the strength of pairing interactions between fermions by a magnetic field. This opens up an exciting possibility to explore  fundamentally new aspects of the nonequilibrium  pairing    following an abrupt change of the coupling strength.  
 
 Interestingly, the dynamical pairing in this regime can be linked  to a seemingly unrelated spin model. The latter describes the interaction of a localized (central) spin with a ``spin bath'' of environmental spins.  The central spin model has reemerged recently in the context of experiments \cite{tarucha,single} on electron spin dynamics  due to the hyperfine interaction with nuclei in GaAs quantum dots. Since single electron quantum dots are now experimentally accessible\cite{single}, and are considered one of the most promising candidates for solid state qubits, it is also important to understand the effect of this interaction on the coherence of the electron spin.

In this paper, we present a theory that provides an accurate description for a range of phenomena in the dynamical pairing and central spin problems. As we will see below, there is an intimate connection between these two problems that will enable us to treat them on an equal footing. 
 
The study of the dynamics of the superconducting state in metals founded on microscopic  principles has begun during the decade following the advent\cite{BCS} of the BCS theory (see Ref.~\onlinecite{kopnin} for a review). The simplest theory for nonstationary processes in superconductors is based on the time-dependent Ginzburg-Landau (TDGL) equation \cite{elihu,schmid,gorkov} for the order parameter $\Delta(t)$. The TDGL approach is valid when quasiparticles are able to reach a local equilibrium quickly on the characteristic time scale of the order parameter variation $\tau_\Delta$ ($\simeq \Delta^{-1}$ at zero temperature). This requirement usually limits the applicability of the TDGL theory to situations where mechanisms destroying Cooper pairs are effective, such as a narrow vicinity of $T_c$ or a large concentration of magnetic impurities. 

An alternative theory \cite{ber,aronov} is the Boltzmann kinetic approach, which describes the dynamics in terms of a kinetic equation for the quasiparticle distribution function coupled to a self-consistent equation for $\Delta(t)$. However, this scheme is justified only when  external parameters change slowly 
on the $\tau_\Delta$ time scale.

As pointed out in Ref.~\onlinecite{barankov}, cold fermionic gases can be in a regime where the notion of the excitation  spectrum is irrelevant and neither TDGL nor the Boltzmann kinetic equations are valid. Indeed, in these systems external parameters such as the position of a Feshbach resonance can change on a time scale $\tau_0\ll\tau_\Delta, \tau_\eps$, where $\tau_\eps$ is the quasiparticle energy relaxation time. On the other hand, the energy relaxation is slow, while the lifetime of the samples is limited. It is therefore desirable to develop a theory that describes nonstationary pairing effects in this regime.

The nonequilibrium Cooper pairing at times $t\ll \tau_\eps$ is non-dissipative  and in a  translationally invariant system can be described by the reduced BCS model
\beg
\hat H_{BCS}={\sum_{{\bf p},\sigma}\eps_{\bf p} \hat c_{{\bf p} \sigma }^\dagger \hat c_{{\bf p} \sigma}- g\sum_{{\bf p},
{\bf k}} \hat c_{{\bf p}\up}^\dagger \hat c_{-{\bf p}\dn}^\dagger \hat c_{-{\bf k}\dn} \hat c_{{\bf k}\up}}
\label{bcsmomentum}
\en
It is interesting to note that the use of this model for describing  the dynamics of a homogeneous superconducting state in metals in the  non-dissipative regime, $t\ll\tau_\eps$, has been justified \cite{volkov} quite generally starting from nonstationary Eliashberg equations.

Here, we are interested in a situation when a system at zero temperature  is out of an equilibrium at $t=0$. The goal is to determine the subsequent evolution of  the initial state. In particular, this includes the case when at $t=0$  the coupling constant has been abruptly changed from $g'$ to $g$.

In the absence of translational invariance, e.g. in a dirty superconductor or in a finite system, Hamiltonian \re{bcsmomentum} has to be appropriately generalized\cite{ander2}.   First, let us discuss the familiar case of electron-electron interactions. If there are no spatial symmetries, but the time-reversal invariance is preserved, single-particle orbitals $\eps_j$ are degenerate only with respect to spin. For each orbital $\eps_j$ there is a pair of states, $|j\up\rangle$ and $|j\dn\rangle$, related by  time reversal symmetry. The pairing occurs between  time reversed states, i.e.
\beg
\hat H_{BCS}={\sum_{j,\sigma}\eps_{j} \hat c_{j \sigma }^\dagger \hat c_{j \sigma}-g\sum_{j, q} \hat c_{j\up}^\dagger \hat c_{j\dn}^\dagger \hat c_{ q\dn} \hat c_{ q\up}}
\label{bcs1}
\en
Deviations of the coupling strength $g$ from a constant in $q$ and $j$ are small and sample-dependent\cite{ander2,kaa}.

For a two species Fermi system, such as a two-component mixture of fermionic atoms in a trap, spin up and down states in \re{bcs1} have to be identified with the two species of fermions. With this replacement, arguments\cite{ander2,kaa, argum} leading to Hamiltonian \re{bcs1} are valid in the weak coupling regime.

Hamiltonian \re{bcsmomentum} is obtained from \re{bcs1} by specializing  to a translationally invariant situation. In this case, time reversed states are $|{\bf p}\up\rangle$ and $|-{\bf p}\dn\rangle$. One can also take a thermodynamic limit by taking the number of orbital levels to infinity. However, in view of applications where the effective number of levels is finite and even small (see below), we deal with the general Hamiltonian \re{bcs1} here.

Remarkably, the  BCS model turns out to be closely related to a model describing the Heisenberg exchange interaction of a single localized spin (the central spin) with a number of  environmental spins. The Hamiltonian reads
\beg
\hat H_0=\sum_{j=1}^{n-1} \gamma_j \hat {\bf K}_0\cdot \hat {\bf K}_j-\B \hat K_0^z
\label{cs1}
\en
where $\hat {\bf K}_0$ is the central spin, $\hat {\bf K}_j$ are environmental spins, $\gamma_j$ are (nonuniform) coupling constants, and $\B$ represents the magnetic field \cite{field}.

 The central spin model has an interesting history of its own. For example, it  emerged in the studies of electron spin dynamics in disordered insulators \cite{subir} and of coherent spin tunnelling in ferromagnetic grains\cite{stamp}.  More recently, it attracted considerable attention as a model for decoherence of qubits \cite{attention,nazarov1,koka,lukin}. In particular, it can be used to model\cite{attention, nazarov1} the hyperfine interaction of a localized electron spin with nuclear spins and the spin-dependent transport\cite{nazarov} in GaAs quantum dots. In these cases, it provides an adequate description of spin dynamics at times shorter than the nuclear spin relaxation time in the regime where the orbital level spacing in the dot is much larger than the temperature and typical interaction energies.

The main result of this paper is an explicit general solution for the dynamics of the BCS and central spin models. In particular, we determine as functions of time the order parameter $\Delta(t)$ and the expectation value of the central spin (equations \re{Jt} and \re{centr}) as well as the dynamics of the remaining degrees of freedom for arbitrary initial conditions. Here, we concentrate on presenting the solution and describing its general properties.  We also mention several applications to specific issues such as nonequilibrium pairing and that  of decoherence due to the hyperfine interaction, but leave a detailed  discussion for the future.

The solution is based on the integrability \cite{rich, gaudin, integr} of the BCS and central spin models. This important property has been largely underestimated in part because it was discovered outside the main physical context of these models. Besides, due to the infinite range of interactions in Hamiltonians \re{bcs1} and \re{cs1}, the mean field approximation is  exact \cite{ander1,largeN} for these models in the limit of large number of particles.
In this approximation, the BCS and central spin models can be mapped onto a classical nonlinear system (see below). However, while the mean field simplifies the description of equilibrium properties to an extent where no advanced techniques are required,  it is the integrability of the resulting classical system that enables us to solve dynamical problems. The BCS solution\cite{BCS} for the ground state and excitations of the BCS model is recovered as a periodic case of the general solution (see the discussion following \re{bcs} and Section~\ref{degenerate}).
It is also interesting to note that, as we will see,  models \re{bcs1} and \re{cs1}, being in many respects classical, have distinct robust features that are preserved when the integrability is destroyed.

The Dicke model \cite{dicke} that describes an ensemble of two-level systems coupled to a bosonic mode is also within the scope of our construction. In fact, it  belongs to the same class \cite{gaudin} of integrable models as the BCS and central spin models. The Dicke model has been recently adopted \cite{andreev} to the study of a dynamical coexistence of atoms and molecules in cold Fermi gases. In the strong coupling regime, it seems to be an appropriate alternative to the BCS model for these systems.  

The general nonstationary solution of the BCS and central spin models is in terms of hyperelliptic functions~-- multiple variable generalizations of ordinary elliptic functions.
These functions frequently arise as solutions of integrable equations and have well known analytical properties
\cite{theta}.

We show that for most initial conditions the dynamics is {\it quasi}-periodic with a number of independent frequencies equal to the number of degrees of freedom, $n$, and evaluate the frequencies in terms of the integrals of motion.
The typical motion  uniformly explores an invariant torus -- an $n$-dimensional subspace of the $2n$-dimensional phase space allowed by the conservation laws. These features of the typical motion are stable against small perturbations that destroy integrability.

Further, we identify certain special values of integrals of motion for which the number of independent frequencies, $m$, becomes less than the number of degrees of freedom. For these degenerate cases, we were able to explicitly reduce the motion of the BCS and central spin models with $n$ degrees of freedom to that of same systems with only $m$ degrees of freedom.  The solution progressively simplifies as the number of independent frequencies decreases. For example, solutions characterized by a single frequency ($m=1$), i.e. periodic trajectories, are given by trigonometric functions; degenerate solutions with two independent frequencies ($m=2$) are given by a combination of trigonometric and elliptic functions (see also Refs.~\onlinecite{barankov, andreev} and the end of Section~\ref{degenerate}) etc.

Periodic solutions ($m=1$) occupy a special place among degenerate solutions. As we show in Section~\ref{degenerate},  they reproduce the BCS solution for the ground state and excitations of the BCS model.  

However, it should be emphasized that, unlike the general solution, degenerate solutions are nonrepresentative of the {\it dynamics} of the BCS and central spin models and, with the exception of  the ground state, are expected  to be unstable.

The paper is organized as follows. In Section~\ref{meanfield}, we use a mean field approximation to map the BCS and central spin models onto equivalent classical models. Section~\ref{solution} contains an explicit general solution for the dynamics of both  problems. In Section~\ref{degenerate}, we consider  degenerate solutions. Section~\ref{open} discusses some open problems and possible applications of our results.

\section{Mean field approximation}
\label{meanfield}

Our starting point is the mean field approximation, which enables the mapping of the  BCS and central spin models onto equivalent classical nonlinear systems. We also derive the equations of motion and describe the relationship between the BCS and central spin models.

The discussion of the mean field is facilitated by representing BCS Hamiltonian \re{bcs1} in terms of Anderson pseudospin-1/2 operators \cite{ander1}.
\beg
\hat H_{BCS}=\sum_{j=0}^{n-1} 2\eps_j \hat K_j^z-g\phantom{.} \hat L_+\hat L_-\quad   \hat {\bf\phantom{,} L}=\sum_{q=0}^{n-1} \hat {\bf K}_q
\label{bcs2}
\en
where $n$ is the number of single-particle orbitals. Pseudospin operators are related to fermion creation and annihilation operators via
\beg
\begin{array}{l}
\dis \hat K_j^z=\frac{c_{j\up}^\dagger \hat c_{j\up}+\hat c_{j\dn}^\dagger \hat c_{j\dn}-1}{2}\\
\\
\dis \hat K_j^-=\hat c_{j\dn} \hat c_{j\up}\quad \hat K_j^+=\hat c^\dagger_{j\up} \hat c^\dagger_{j\dn},\\
\end{array}
\label{pseudo}
\en 
Pseudospins are defined on unoccupied and doubly occupied pairs of  states $|j\up\rangle$ and $|j\dn\rangle$,
where they have all properties of  spin-1/2. Singly occupied pairs of states, on the other hand, do not participate in pair scattering and are decoupled from the dynamics.

Our goal is to determine the time evolution according to Hamiltonian \re{bcs2} of an arbitrary initial distribution of pseudospins.
The mean field approximation consists in the replacement of the effective field seen by each pseudospin in BCS Hamiltonian \re{bcs2} with its quantum mechanical average, ${\bf b}_j(t)=(-2\Delta_x(t), -2\Delta_y(t), 2\eps_j)$, where $\Delta(t)$ is the BCS gap function
\beg
\Delta(t)\equiv\Delta_x(t)-i\Delta_y(t)\equiv g\sum_p \langle \hat K_j^-(t)\rangle
\label{gap}
\en
In this approximation, each spin evolves in the self-consistent field created by other spins
\beg
\dot {\hat {\bf K}}_j=i\left[ \hat H_{BCS}, \hat {\bf K}_j \right]={\bf b}_j \times \hat {\bf K}_j
\label{spins1}
\en
Since equations \re{spins1} are linear in $\hat {\bf K}_j$, we can take their quantum mechanical average with respect to the time-dependent state of the system to obtain
\beg
\dot {\bf s}_j={\bf b}_j \times {\bf s}_j \phantom{m} {\bf b}_j=\left(-gJ_x, -g J_y, 2\eps_j\right)\quad {\bf J}=\sum_{q=0}^{n-1} {\bf s}_q
\label{spins2}
\en
where ${\bf s}_j(t)=2\langle \hat {\bf K}_j(t) \rangle$. Evolution equations \re{spins2} conserve the square of the average for each spin, i.e. ${\bf s}_j^2={\rm const}$. If spins initially were in a product state \cite{product}, ${\bf s}_j^2=1$. 

Note from equation \re{pseudo} the following correspondence between components of  ${\bf s}_j(t)$ and the normal and anomalous (Keldysh) Green functions at coinciding times:
$$
\begin{array}{l}
\dis G_j(t)=-i\langle[ c_{j\up}(t),  \hat c_{j\up}^\dagger(t)]\rangle=is_j^z(t)\\
\\
\dis F_j(t)=-i\langle[ c_{j\up}(t),  \hat c_{j\dn}(t)]\rangle=is_j^-(t)\\
\end{array}
$$
Equations \re{spins2} were derived \cite{volkov} for phonon superconductors  within the general framework of nonstationary Eliashberg theory in the collisionless regime $t\ll\tau_\eps$. A linearized version of these equations was considered  in Refs~\onlinecite{ander1, volkov} and \onlinecite{galperin}.  

Due to  the infinite range of interactions between spins in \re{bcs2}, the mean field approximation is  exact  in the thermodynamic limit. 
  For a system with a finite number, $N$, of particles (spins) we expect, based on an analysis of  leading finite size corrections \cite{fs} to the mean field, equations \re{spins2} to be accurate for large $N$ at times $t<t^* N^\eta$, where $t^*$ and $\eta>0$ do not depend on $N$.

We see that ${\bf s}_j=2\langle \hat {\bf K}_j \rangle$ have all properties of classical spins governed by a Hamiltonian 
\beg
H_{BCS}=\sum_{j=0}^{n-1} 2\eps_j s_j^z-\frac{g}{2} J_+J_-
\label{bcs}
\en
and usual angular momentum Poisson brackets. Thus, the problem reduces to  determining the time evolution of $n$ classical spins according to Hamiltonian \re{bcs}.

The BCS solution for the ground state corresponds to the minimum of \re{bcs} and is obtained  by aligning each spin in \re{bcs} antiparallel to the field acting upon it. A pair excitation\cite{pair} (excitation of the condensate) of energy $2\sqrt{(\eps_j-\mu)^2+\Delta_0^2}$, where $\mu$ and $\Delta_0$ are the chemical potential and the equilibrium gap respectively,  is obtained by flipping the spin ${\bf s}_j$.   The $U(1)$ symmetry of the BCS order parameter $\Delta$
is equivalent to the symmetry of \re{bcs} with respect to uniform rotations of all spins around the $z$-axis.
Due to this symmetry, spin configurations that determine the ground state and excitations can rotate around the $z$-axis with a frequency $2\mu$ at no energy cost.  
In the presence of a particle-hole symmetry  $\mu=0$ and these  configurations are stationary with respect to Hamiltonian \re{bcs} (see Section~\ref{degenerate} where we recover the BCS solution as a periodic case of the general time-dependent solution).

Next, we turn to the central spin model and its relationship to the BCS model.
The central spin Hamiltonian \re{cs1} turns out to be a member of an integrable family of Gaudin magnets \cite{gaudin} -- $n$ Hamiltonians of the form:
\beg
\hat H_q=2\sum_{j=0}^{n-1}{\lefteqn{\phantom{\sum}}}' \phantom{.}\frac{ \hat {\bf K}_q\cdot \hat {\bf K}_j}{\eps_q-\eps_j}-\alpha \hat K_q^z\quad q=0,\dots,n-1
\label{gaudin1}
\en
where $\eps_j$ and $\alpha$ are arbitrary parameters.  All Hamiltonians $\hat H_q$ commute with each other and with the $z$-component of the total spin, $\hat L_z\propto\sum_q \hat H_q$, for arbitrary $\eps_j$ and $\alpha$.
The central spin model \re{cs1} coincides with $\hat H_0$ in equation \re{gaudin1} if we choose
$\eps_0-\eps_j=2/\gamma_j$ and $\alpha=\B$.

Note that if a number of orbitals $\eps_j$ in \re{bcs2} are degenerate, the magnitude of their total spin $\sum_{\eps_j=\mbox{\scriptsize const}} \hat {\bf K}_j$ is conserved by Hamiltonian \re{bcs2}. In this case one can replace ${\hat {\bf K}_j\to\sum_{\eps_j=\mbox{\scriptsize const}} \hat {\bf K}_j}$ in Hamiltonians \re{bcs2} and \re{gaudin1} and sum over nondegenerate orbitals only.

Now consider the following linear combination of $\hat H_q$:
$$
  \sum_{q=0}^{n-1}\eps_q \hat H_q=-\frac{\alpha}{2}\left[\sum_{q=0}^{n-1} 2\eps_q \hat K_q^z-\frac{2}{\alpha}\hat L_+
\hat L_-\right]+\mbox{const}
$$
We see that for $\alpha=2/{g}$ the expression in the square brackets  coincides with  BCS Hamiltonian \re{bcs2}. Therefore, for $\alpha=2/{g}$, the BCS Hamiltonian commutes with all $\hat H_q$  and thus belongs to the same class of integrable models \cite{gaudin, integr}.

In the same way as we did for the BCS model, we employ the mean field approximation to derive classical Hamiltonians that govern the evolution of ${{\bf s}_j(t)=2\langle \hat {\bf K}_j(t) \rangle}$ for Hamiltonians \re{gaudin1}:  
\beg
\left\{
\begin{array}{l}
\dis H_q=\sum_{j=0}^{n-1}{\lefteqn{\phantom{\sum}}}' \phantom{.}\frac{ {\bf s}_q\cdot{\bf s}_j}{\eps_q-\eps_j}-\alpha s_q^z\\
\\
\dis H_{BCS}=\sum_{j=0}^{n-1} 2\eps_j s_j^z-\frac{ J_+J_-}{\alpha}\\
\end{array}\right.
\label{class}
\en
In particular, the classical counterpart of  central spin Hamiltonian \re{cs1} is
\beg
H_0=\sum_{j=0}^{n-1} \frac{\gamma_j}{2} {\bf s}_0\cdot{\bf s}_j-\B s_0^z
\label{csclass}
\en
The validity of the mean field approximation for  central spin model \re{cs1} is subject to similar considerations as for the BCS model (see the second paragraph following equation \re{spins2} and also Ref.~\onlinecite{nazarov1}).  

Thus,  the problem of determining the dynamics of the BCS and central spin models reduces to solving equations of motion for  Hamiltonians $H_0$ and $H_{BCS}$ in system  \re{class}. 
 To obtain solutions for  BCS \re{bcs1} and central spin \re{cs1} models, one has to choose
\beg
\begin{array}{l}
\dis \alpha=\frac{2}{g}\quad \mbox{BCS}\\
\\
 \dis \frac{2}{\eps_0-\eps_j}=\gamma_j \quad \alpha=\B\quad \mbox{central spin}
 \end{array}
 \label{choice}
 \en

Hamiltonians \re{class} Poisson-commute with each other and with the $z$-component of the total spin $J_z\propto \sum_q H_q$. Each Hamiltonian in system \re{class} describes an evolution of $n$ spins in an $2n$-dimensional phase space (two angles for each spin) and has $n$ integrals of motion (the energy and the remaining Hamiltonians from system \re{class}). 
Therefore, all Hamiltonians \re{class} are classical integrable models.

\section{The solution}
\label{solution}

In the previous section we mapped  the BCS and central spin models onto a classical integrable system \re{class}.  Nevertheless, even though Liouville's theorem \cite{arnold} guarantees a formal integrability in quadratures of classical integrable models, an explicit solution for the evolution of the original dynamical variables is not always possible. Fortunately, this turns out to be not the case for the BCS and central spin problems.   For these models, as detailed below, we were able to obtain an explicit general solution of equations of motion.  

Before we proceed, let us discuss  generic features \cite{arnold}  expected of the dynamics of a classical integrable model with $n$ degrees of freedom. The motion  is confined by  conservation laws to an $n$-dimensional subspace of the $2n$-dimensional phase space. This subspace (invariant torus) is determined by initial values of integrals of motion and is topologically equivalent to an $n$-dimensional torus.  The motion on the invariant torus is characterized by $n$ angle variables, $\phi_k$ and the corresponding angular frequencies, $\phi_k(t)=\Omega_k t+\phi_k(0)$. The frequencies depend only on integrals of motion, while constants $\phi_k(0)$ are determined by initial values of remaining degrees of freedom. Since for most initial conditions all frequencies $\Omega_k$ are independent\cite{indep},  typical trajectories uniformly explore the entire torus.  All these properties are not affected by small perturbations destroying integrability (Kolmogorov-Arnold-Moser theorem). 

Now we turn to the solution for the dynamics of the BCS and central spin models. The solution consists of two main steps. The first one is a change of variables that casts  equations of motion into  the form of a known mathematical problem.  In the second step, we use the solution of this problem to obtain  dynamical variables ${\bf s}_j(t)=2\langle \hat {\bf K}_j(t) \rangle$ as explicit functions of time.

  The change of variables is facilitated by defining a $2\times 2$ matrix\cite{lax} that depends on dynamical variables ${\bf s}_j$ and also on an auxiliary parameter $u$ 
\beg
{\cal L}=\left(
\begin{array}{ll}
A(u) & \phantom{-}B(u)\\
 B^*(u) & -A(u)\\
 \end{array}
 \right)
 \label{lax}
 \en
 where  matrix elements of ${\cal L}$ are
 \beg
 A(u)=\alpha-\sum_{j=0}^{n-1} \frac{s_j^z}{u-\eps_j}\quad B(u)=\sum_{j=0}^{n-1}\frac{s_j^-}{u-\eps_j}
 \label{AB}
\en
The eigenvalues of this matrix, $\pm v(u)$, depend on dynamical variables  only through integrals of motion -- $H_j$ 
and~${\bf s}^2_j$
\beg
v^2(u)=\alpha^2+\sum_{j=0}^{n-1} \left( \frac{2 H_j}{u-\eps_j}+\frac{{\bf s}^2_j}{(u-\eps_j)^2}\right)\equiv\frac{\alpha^2 Q_{2n}(u)}{P_n^2(u)}
\label{v}
\en
where we defined two polynomials, $Q_{2n}(u)$ and $P_n(u)$ that will be frequently used in subsequent calculations
\beg
Q_{2n}(u)\equiv\prod_{j=0}^{2n-1} (u-E_j)\equiv\sum_{k=0}^{2n} \alpha_k u^k\quad \alpha_{2n}=1
\label{Q}
\en
\beg
 P_n(u)\equiv\prod_{q=0}^{n-1}(u-\eps_q) 
 \label{P}
\en
Note  that
because ${\cal L}$ is Hermitian for real $u$, its eigenvalues are real  and therefore the polynomial $Q_{2n}(u)$ is positive definite on the real axis. We will often refer to the polynomial $Q_{2n}(u)$  as the spectral polynomial.

Following an algebraic version of variable separation  method,\cite{sklyanin, vadim} we introduce $n-1$ variables, $u_j$, as zeroes of $B(u)$ -- one of the off-diagonal matrix elements of ${\cal L}$. Variables $v_j$, canonically conjugate to $u_j$, are given by one of the eigenvalues of matrix ${\cal L}$ at $u=u_j$.
\beg
B(u_j)=0\quad v_j=-A(u_j)\quad j=1,\dots, n-1
\label{u}
\en
Although we will not use the Hamilton-Jacobi method, we remark that variables $(u_j, v_j)$ separate\cite{vadim} Hamilton-Jacobi equations for Hamiltonians \re{class}. 
Since $u_j$ are zeroes of $B(u)$, we can rewrite the matrix element $B(u)$ as
\beg
B(u)=J_-\frac{\prod_{j=1}^{n-1}(u-u_j)}{\prod_{j=0}^{n-1} (u-\eps_j)}\equiv J_-\frac{R_{n-1}(u)}{P_n(u)} 
\label{B}
\en
This equation is useful for expressing original dynamical variables -- components of spins ${\bf s}_j$ -- through separation variables $u_j$ and $J_-$. For example, using equations \re{AB} and \re{B}, we have
\beg
s_j^-=\mathop{\mbox{res}}_{u=\eps_j} B(u)= J_-\frac{R_{n-1}(\eps_j)}{P'_n(\eps_j)}
\label{res}
\en
where the prime denotes a derivative with respect to $u$.

Equations of motion for  central spin Hamiltonian \re{csclass} in terms of new variables $u_j$ and $J_-$ can be derived by evaluating Poisson brackets between the  Hamiltonian and  matrix elements of ${\cal L}$.
We have
\beg
\begin{array}{l}
\dis \dot u_j=\frac{i \B\sqrt{Q_{2n}(u_j)}}{u_j\prod_{m\ne j}(u_j-u_m)} \frac{s_0^-}{J_-}\\
\\
 \dis \dot J_-=i\B s_0^-=i\lambda  J_-R_{n-1}(0)\quad \lambda=\frac{\B}{2^n} \prod_{j=1}^n \gamma_j\\
 \end{array}
\label{h0ev}
\en
Similarly, one obtains  equations of motion for  BCS Hamiltonian \re{bcs}
\beg
\begin{array}{l}
\dis \dot u_j=\frac{2i \sqrt{Q_{2n}(u_j)}}{\prod_{m\ne j}(u_j-u_m)} \\
\\
 \dis  \dot J_-=-2i J_-\left(\sum_{j=0}^{n-1} \eps_j+\frac{g J_z}{2}-\sum_{j=1}^{n-1} u_j\right)\\
 \end{array}
\label{bcsev}
\en
In equations  \re{h0ev} and \re{bcsev} as well as in the rest of the paper, with no loss of generality, we shifted parameters $\eps_j$ in the BCS and central spin Hamiltonians by a constant so that $\eps_0=0$. 

In the reminder of this section we obtain an explicit general solution for the expectation value of each (pseudo)spin, ${\bf s}_j(t)=2\langle \hat {\bf K}_j(t) \rangle$, as a function of time. In particular, this includes the expectation value of the central spin, ${\bf s}_0(t)$, and the BCS gap function, ${\Delta(t)=g J_-(t)}$.

The solution is based on the observation that, with a help of elementary algebra, equations of motion \re{h0ev} and \re{bcsev} can be cast into a form of a known mathematical problem. Specifically, one can rewrite equations  \re{h0ev} and \re{bcsev} for variables $u_j$ as
\beg
\sum_{j=1}^{n-1} \frac{u_j^{l-1} du_j}{\sqrt{Q_{2n}(u_j)}}=d{x}_l\quad l=1,\dots, n-1
\label{jip}
\en
where the polynomial $Q_{2n}(u)$ is defined in equation \re{v}.
To obtain equations of motion \re{bcsev} for the BCS Hamiltonian  one has to choose
\beg
{\bf x}^T=i(c_1,\dots, c_{n-2}, 2t+c_{n-1})
\label{wbcs}
\en
while for  the central spin model one has
\beg
{\bf x}^T=-i\lambda(t+c_1,\dots, c_{n-2}, c_{n-1})
\label{wcs}
\en
where $c_1,\dots,c_{n-1}$ are arbitrary constants.

Differential equations \re{jip} constitute a well-known mathematical problem called  Jacobi's inversion problem (see, for example, Ref.~\onlinecite{theta} and references therein). The solution of equations \re{jip} can be expressed \cite{victor} through hyperelliptic Abelian functions -- multiple variable generalizations of ordinary elliptic functions.
These functions are often encountered as solutions of integrable equations and have well known analytical properties. They are also implemented in standard mathematical software packages \cite{maple}.

Riemann theta function of {\it genus} $\g$ (in our case ${\g=n-1}$) is defined as the following  sum over all $\g$-dimensional integer vectors
\beg
\theta({\bf x}|\tau)=\sum_{{\bf m}\in Z^{\scriptsize \g} } \exp\left[ i\pi({\bf m}^T\tau{\bf m}+2{\bf x}^T {\bf m})\right]
\label{theta}
\en
where $\tau=\omega'\omega^{-1}$, and $\omega$ and $\omega'$ are $\g\times \g$ matrices to be specified below.
Klenian  $\sigma$- and $\zeta$-functions of genus $\g$ are defined through the Riemann theta function as\cite{rimconst}
\beg
\begin{array}{l}
\sigma({\bf x})=C \exp[ {\bf x}^T\eta (2\omega)^{-1}{\bf x}] \theta\left((2\omega)^{-1}{\bf x}|\tau\right)\\
\\
\dis \zeta_l({\bf x})=\frac{\partial \ln\sigma({\bf x}) }{\partial x_l}\\
\end{array}
\label{sigma}
\en

In our case, the $\g\times \g$ matrices $\omega$ and $\eta$ (matrices of periods) that appear in the definition of hyperelliptic functions (\ref{theta}) and \re{sigma} are 
\beg
\begin{array}{l}
\dis 2\omega_{kl}=\oint_{b_k} \frac{u^{l-1} du}{\sqrt{Q_{2n}(u)}}\\
\\
\dis 2\eta_{kl}= -\oint_{b_k} \frac{du}{4\sqrt{Q_{2n}(u)}} \sum_{k=l+1}^{2n-l} (k-l)\alpha_{k+l}u^{k-1}\\
 \end{array}
\label{periods}
\en
where  contours of integration $b_k$ go around branch cuts of a hyperelliptic {\it curve} $z(u)=\sqrt{Q_{2n}(u)}$  as shown on Fig.~1.
Matrices $\omega'$ and $\eta'$ are also defined by equations \re{periods} with the replacement of the contours of integration $b_k\to h_k$ (see Fig.~1).

\begin{figure}[ht]
\epsfysize=6.5cm 
\vspace{2.4cm}
\centerline{\epsfbox{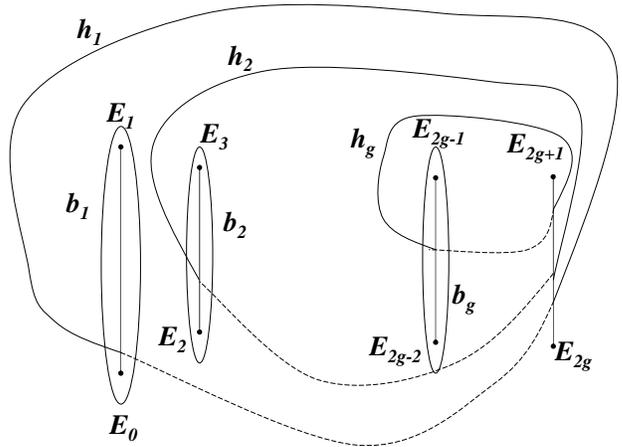}}
\vspace{-3.5cm}
\caption{Riemann surface of a hyperelliptic curve $z(u)=\sqrt{Q_{2n}(u)}$ of genus\cite{genus} $\g=n-1$ showing contours of integration $b_k$ and $h_k$ with $k=1,\dots,\g$. These contours appear in the definition of matrices of periods \re{periods} of hyperelliptic functions. Contours $b_k$ go around $\g=n-1$ branch cuts connecting first $2\g=2n-2$ roots $E_k$ of the spectral polynomial $Q_{2n}(u)$. Integration along the contours is clockwise. Note that since the spectral polynomial is positively defined the branch cuts are parallel to the imaginary axis. }
\end{figure}

The solution \cite{victor} of differential equations \re{jip} for arbitrary initial conditions  states that functions $u_p({\bf x})$ are  zeroes of the following  polynomial of a degree $\g=n-1$:
\beg
R_{n-1}(u,t)=u^{n-1}-\sum_{k=1}^{n-1} f_k({\bf x}) u^{k-1}
\label{R}
\en
where coefficients of $R_{n-1}(u, t)$ are
\beg
f_k({\bf x})=\zeta_k({\bf x}+{\bf d})- \zeta_k({\bf x}-{\bf d})+a_k
\label{f}
\en
Here  vector ${\bf x}$ is defined by equations \re{wbcs} and \re{wcs} for the BCS and central spin problems respectively. 
The argument of $\zeta$-functions in equations \re{f} is shifted by a vector of constants ${\bf d}$ that has the following components:
\beg
d_l=\int_{E_0}^\infty \frac{ u^{l-1} du}{\sqrt{Q_{2n}(u)}}\quad l=1,\dots,n-1
\label{d}
\en
where $E_0$ is one of the roots of the spectral polynomial $Q_{2n}(u)$ (see Fig.~1 and equation \re{Q}). Note that, according to its definition \re{v}, the spectral polynomial $Q_{2n}(u)$ depends only on integrals of motion and on parameters $\eps_j$ and $\alpha$.

Finally, constants $a_k$ in equations \re{f} are obtained from the expansion
\beg
\sqrt{\sum_{k=0}^{2n} \frac{\alpha_k}{z^k}}=\frac{1}{z^n}+\frac{a_1}{z^{n-1}}+\dots+\frac{a_{n-1}}{z}+O(1)
\label{a}
\en
where $\alpha_k$ are  coefficients of  the spectral polynomial $Q_{2n}(u)$ defined by equation \re{Q}. For example, $a_{n-1}=\alpha_{2n-1}/2$, $a_{n-2}=\alpha_{2n-2}/2-\alpha^2_{2n-1}/8$ etc.
 
To obtain quantum mechanical expectation values of original spins ${\bf s}_j(t)=2\langle \hat {\bf K}_j(t) \rangle$ for  central spin model \re{cs1}  and pseudospins \re{pseudo}  for  BCS model \re{bcs1}, we observe that, according to equation \re{res},  they can be expressed through the polynomial $R_{n-1}(u, t)$ given by expression \re{R}. We have
\beg
 s_j^-(t)=J_-(t)\frac{R_{n-1}(\eps_j, t)}{P'_n(\eps_j)}
\label{trans}
\en
Recall that coefficients of the polynomial $P_n(u)$, defined by equation \re{P}, depend only on parameters $\eps_j$.

Similarly, one can  use an expression for the matrix element $A(u)$ defined in equation \re{AB} in terms of variables $u_j$ to obtain $z$-components of (pseudo)spins as functions of time.  However, these components are not independent and can be also obtained from  the equation  $(s_j^z)^2+|s_j^-|^2={\bf s}_j^2=1$, where the sign of $s_j^z$ is determined by initial conditions. 

Integrating equations \re{h0ev} and \re{bcsev} for the evolution of $x$ and $y$ components of the total (pseudo)spin with the help of equation \re{f},  we obtain
\beg
 J_-(t)=c_n e^{-i\beta t}\frac{\sigma({\bf x}+{\bf d})}{\sigma({\bf x}-{\bf d})}\quad \Delta(t)=g J_-(t)
\label{Jt}
\en
where vector ${\bf x}$ is defined by equations (\ref{wbcs}) and (\ref{wcs}), the frequency $\beta$ is different for the BCS ($\beta_{BCS}$) and central spin ($\beta_0$) models
\beg
\beta_{BCS}=g J_z+2\sum_{j=0}^{n-1}\eps_j-\alpha_{2n-1}\quad \beta_0=a_1\lambda
\label{beta}
\en
and $\lambda$ is defined in equation \re{h0ev}.

Average components of the electron (central) spin, ${\bf s}_0(t)=2\langle \hat {\bf K}_0(t)\rangle$ are obtained from equation \re{trans} (recall that $\eps_0=0$)
\beg
s_0^-(t)= c'_n\left[\zeta_1({\bf x}-{\bf d})-\zeta_1({\bf x}+{\bf d}) -a_1\right]\frac{\sigma({\bf x}+{\bf d})}{\sigma({\bf x}-{\bf d})}
\label{centr}
\en
where $c'_n=(\lam c_n)/\B$.

To summarize, quantum mechanical averages of (pseudo)spins for the BCS and central spin models, including the electron (central) spin and the BCS gap function $\Delta(t)$  are hyperelliptic functions (equations \re{theta} and \re{sigma}) of the time variable and $n-1$ constants $c_k$ (equations \re{trans}, \re{Jt}, and \re{centr}).  These functions are specified by their matrices of periods $\omega$, $\omega'$, and $\eta$. The latter depend only on integrals of motion according to equations \re{v} and \re{periods}.  The remaining $n$ out of $2n$ initial conditions fix constants $c_k$ in equations (\ref{wbcs}) and (\ref{wcs}).

An important characteristic of a quasi-periodic motion are its (Fourier) frequencies. To determine the frequencies for the  BCS and central spin models, we observe from equation \re{sigma} that the
time variable enters the solution only through combinations of $e^{i\Omega_k t}$, where
\beg
\begin{array}{l}
\Omega_k=i\pi \omega^{-1}_{k{\scriptsize \g}}\quad \Omega_0=\beta_{BCS}\quad \mbox{BCS}\\
\\
\Omega_k=-i\pi \omega^{-1}_{k1}\quad \Omega_0=\beta_{0}\quad \mbox{central spin}\\
\end{array}
\label{freq}
\en
 $k=1,\dots,\g$, and $\g=n-1$.  Note that, since the spectral polynomial $Q_{2n}(u)$ in equation \re{periods} is positively defined, the matrix $\omega$ is pure imaginary and, therefore, all frequencies $\Omega_k$ are real (see Fig.~1).

\section{Degenerate solutions}
\label{degenerate}

Here we identify a class of ``resonant'' nonstationary solutions for  the BCS and central spin models with $n$ degrees of freedom ($n$ spins or pseudospins) but only $m<n$ independent frequencies. 

While a typical dynamics of the BCS and central spin models with $n$ (pseudo)spins is quasi-periodic with $n$ independent frequencies, it is clear already on physical grounds that there must be solutions with fewer frequencies. For example, the only ``spatial'' symmetry of the BCS ground state is the $U(1)$ symmetry of the order parameter that translates into rotations of all pseudospins in Hamiltonians \re{bcs2} and \re{bcs} around the $z$-axis. The corresponding trajectory in the phase space is a circle and, therefore, the motion is periodic with a single frequency (see also the discussion following equation~\re{bcs} and below in this section). 

Generally, one can show\cite{arnold} that there are (resonant) tori with an arbitrary number $m<n$ of independent frequencies in a vicinity of any point in the phase space. It should be noted though that the set of points for which $m<n$ has a zero measure in the phase space just like the set of rational numbers on the real axis. Moreover, such points are typically unstable\cite{arnold} and provide seeds of chaotic behavior in integrable systems.  This occurs because small perturbations are able to destroy resonant tori  and let the corresponding trajectories escape to other regions of the energy shell. In contrast, the majority of invariant tori with $n$ independent frequencies are only slightly deformed by perturbations.

 Interestingly, for the resonant (degenerate) cases considered here we were able to reduce the BCS and central spin models with $n$ spins to  same models with only $m$  spins. For example, solutions  with $m=2$  frequencies can be obtained from  the BCS and central spin models for only two spins etc.
Further, we will see that one spin solutions (periodic trajectories) are special among  degenerate solutions in that they  correspond to the BCS energy spectrum. The solution that corresponds to the ground state is further exceptional in that it is stable against conservative perturbations because it minimizes the total energy. In this case, even though there is only one frequency, the trajectory cannot leave the resonant torus because the latter coincides with the energy shell. The same applies to the solution that minimizes the central spin Hamiltonian.

We start by determining conditions under which the number of independent frequencies is reduced. Since the frequencies are fixed by initial values of integrals of motion, degeneracies occur only for special values of the integrals. On the other hand, a complete information about integrals of motion is encoded in the spectral polynomial $Q_{2n}(u)$ defined in equations \re{v} and \re{Q}.
Specifically, we have seen in the previous section that the number of frequencies $\Omega_k$ is equal to the number of branch cuts of the hyperelliptic   curve $z(u)=\sqrt{Q_{2n}(u)}$.   Evidently, the latter decreases by one when two roots of the polynomial $Q_{2n}(u)$ coincide. Thus, as detailed below, merging $2(n-m)$ roots of $Q_{2n}(u)$ we obtain solutions with $m<n$ frequencies. 

Indeed, consider a situation when $2(n-m)$ roots coincide, i.e. the spectral polynomial $Q_{2n}(u)$ in equation \re{v} has the following special form:
\beg
Q_{2n}(u)=\widetilde{Q}_{2m}(u)\prod_{k=m}^{n-1}(u-E_k)^2=\widetilde{Q}_{2m}(u)W^2_{n-m}(u)
\label{red1}
\en
We assume that  double roots in equation \re{red1} are complex conjugate to each other so that coefficients of 
polynomials $\widetilde{Q}_{2m}(u)$ and $W_{n-m}(u)$ are real. Note  that, since the polynomial $Q_{2n}(u)$ is positively defined (see the remark following equation \re{P}), so is the polynomial  $\widetilde{Q}_{2m}(u)$. We will see below that degenerate solutions are completely determined by the ``residual'' spectral polynomial $\widetilde{Q}_{2m}(u)$.

Now let us  choose $m-n$ separation variables $u_j$ to coincide with double roots of the polynomial $Q_{2n}(u)$, i.e. 
\beg
\begin{array}{l}
\dis u_j=E_j \quad j=m,\dots,n-1\quad\\
\\
 \dis W_{n-m}(u)=\prod_{k=m}^{n-1}(u-E_k)=\prod_{j=m}^{n-1}(u-u_j)\\
 \end{array}
 \label{st}
\en
We observe that equations of motion  for these variables are automatically satisfied because both sides of equations \re{h0ev} and \re{bcsev} vanish.  

 Next, we use equations \re{red1} and \re{v}  to express eigenvalues of matrix ${\cal L}$  in terms of the residual spectral polynomial $\widetilde{Q}_{2m}(u)$ as follows
\beg
v^2(u)=\alpha^2 \biggl ( \sum_{q=0}^{n-1} \frac{C_q}{u-\eps_q}\biggr)^2 \widetilde{Q}_{2m}(u)
\label{red2}
\en
where we expanded the ratio of polynomials $W_{n-m}(u)$ and $P_n(u)$ into elementary fractions
\beg
\frac{W_{n-m}(u)}{P_n(u)}=\sum_{q=0}^{n-1} \frac{C_q}{u-\eps_q}
\label{ratios}
\en
Since the degrees of polynomials $W_{n-m}(u)$ and $P_n(u)$  differ by $m$, there are $m$ constraints on the values of $C_q$, which can be derived by expanding equation \re{ratios} in $1/u$. We have  
\beg
\sum_{q=0}^{n-1} C_q\eps_q^{k-1}=\delta_{km}\quad k=1,\dots,m
\label{cond}
\en
Values of $C_q$  can be expressed through coefficients of $\widetilde{Q}_{2m}(u)$ by comparing residues at  double poles at $u=\eps_q$ in equations \re{v} and \re{red2}. 
\beg
C_q=\frac{e_q}{\alpha \sqrt{\widetilde{Q}_{2m}(\eps_q)} }\quad\mbox{where}\quad e_q=\pm 1
\label{c}
\en
Note that equations \re{cond} provide the following $m$ constraints 
\beg
\sum_{q=0}^{n-1}\frac{e_q\eps_q^{k-1}}{ \sqrt{\widetilde{Q}_{2m}(\eps_q)} }=\alpha\delta_{km}\quad \quad k=1,\dots,m
\label{gap1}
\en
on  $2m$ coefficients of a positively defined polynomial  
$\widetilde{Q}_{2m}(u)$.

Values of integrals of motion $H_q$ for which degeneracies  occur can be determined in terms of coefficients of the residual spectral polynomial $\widetilde{Q}_{2m}(u)$ by comparing residues at simple poles at $u=\eps_q$ in equations \re{v} and \re{red2}. 
\beg
H_q=\alpha^2\sum_{j=0}^{n-1}{\lefteqn{\phantom{\sum}}}' \phantom{.}\frac{ C_q C_j \widetilde{Q}_{2m}(\eps_q)}{\eps_q-\eps_j}+\frac{\alpha^2}{2} C_q^2 \widetilde{Q}'_{2m}(\eps_q) 
\label{values}
\en
In particular, the value of $H_0$ -- twice the energy\cite{energy} of the central spin model on degenerate solutions is
\beg
{\cal E}_0=h\sqrt{\widetilde{\alpha}_0} e_0\sum_{q=1}^{n-1} \frac{C_j\gamma_j}{4}+\frac{\widetilde{\alpha}_1}{2\widetilde{\alpha}_0}
\label{E0}
\en
where $\widetilde{\alpha}_k$ are coefficients of $\widetilde{Q}_{2m}(u)$ defined as
\beg
 \widetilde{Q}_{2m}(u) \equiv\sum_{k=0}^{2m} \widetilde{\alpha}_k u^k \quad \widetilde{\alpha}_{2m}=1
\label{tildeQ}
\en
The BCS energy ${\cal E}_{BCS}\propto \sum_q \eps_qH_q +\mbox{const}$ and the value of $J_z \propto \sum_q H_q$ can be obtained from the knowledge of $H_q$. However, a more convenient derivation is to compare  expansions of equations \re{v} and \re{red2} in powers of $1/u$. We have 
\beg
J_z+\alpha\sum_{q=0}^{n-1} C_q\eps_q^m=-\frac{\alpha \widetilde{\alpha}_{2m-1}}{2}
\label{jz}
\en
\beg
\begin{array}{l}
\dis g {\cal E}_{BCS}=-\sum_{q=0}^{m-1} C_q (2\eps_q^{m+1}+\widetilde{\alpha}_{2m-1}\eps_q^m)-\\
\\
\dis \qquad\qquad\frac{\widetilde{\alpha}_{2m-1}^2}{4}+\widetilde{\alpha}_{2m-2}\\
\end{array}
\label{ebcs}
\en

So far, we determined  initial conditions for which the number of independent frequencies for a system with $n$ (pseudo)spins drops from $n$ to $m$.  We saw that these conditions are fixed by the residual spectral polynomial $\widetilde{Q}_{2m}(u)$. Note that separation variables $u_j$ with $j\ge m$ are also fixed by this polynomial because they are zeroes of equation \re{ratios}. 

Thus, we are left with $m$ dynamical variables -- $u_j$ with $j=1,\dots,m-1$ and $J_-$. Equations of motion \re{h0ev} and \re{bcsev} for these $u_j$ can be cast into a standard form \re{jip} where $n$ is now replaced with $m$ and the original spectral polynomial $Q_{2n}(u)$ is replaced with the residual polynomial $\widetilde{Q}_{2m}(u)$. On the other hand, it is clear that we would obtain similar equations of motion for $m$ spins ${\bf \widetilde{s}}_j$ governed by  the BCS or central spin Hamiltonians. This analogy can be made precise if we identify the residual polynomial $\widetilde{Q}_{2m}(u)$ with the spectral polynomial of a system of $m$ spins (cf. equation \re{v}), i.e. 
\beg
\widetilde{v}^2=\alpha^2+\sum_{j=0}^{m-1} \left( \frac{2 \widetilde{H}_j}{u-\widetilde{\eps}_j}+\frac{{\bf \widetilde{s}}^2_j}{(u-\widetilde{\eps}_j)^2}\right)=\frac{\alpha^2 \widetilde{Q}_{2m}(u)}{\widetilde{P}_m^2(u)}
\label{tildev}
\en
where we replaced parameters $\eps_j$ with new ones 
\beg
\widetilde{P}_m(u)=\prod_{q=0}^{m-1}(u-\widetilde{\eps}_q)
\label{Ptilde}
\en
and chose $\widetilde{\eps}_0=0$, consistent with the choice $\eps_0=0$ (see below equation \re{bcsev}). The rest of $\widetilde{\eps}_j$ are arbitrary real numbers.

BCS \re{bcs} and central spin \re{csclass} Hamiltonians for new classical spins ${\bf \widetilde{s}}_j$ are
\beg
\begin{array}{l}
\dis \widetilde{H}_{BCS}=\sum_{j=0}^{m-1} 2\widetilde{\eps}_j \widetilde{s}_j^z-\frac{g}{2} \widetilde{J}_+\widetilde{J}_- \quad \widetilde{{\bf J}}=\sum_{q=0}^{m-1} \widetilde{{\bf s}}_q\\
\\
\dis \widetilde{H}_0=\sum_{j=0}^{m-1} \frac{\widetilde{\gamma}_j}{2} \widetilde{{\bf s}}_0\cdot{\widetilde{\bf s}}_j-\B \widetilde{s}_0^z\quad \widetilde{\gamma}_j=\frac{2}{\widetilde{\eps}_0-\widetilde{\eps}_j}\\
\end{array}
\label{htilde}
\en

Now, comparing equations of motion \re{h0ev} and \re{bcsev} for $m$ separation variables $\widetilde{u}_j$ for new spins with those for the remaining variables $u_j$ for $j=1,\dots,m-1$ in the original problem with $n$ spins, we see that they coincide if we re-scale the time variable for the central spin problem as follows 
\beg
\begin{array}{ll}
\dis \widetilde{t}=\frac{(-1)^ne_0\prod_{k=1}^{m-1}\widetilde{\eps}_m}{\B\sqrt{\widetilde{\alpha}_0}}\phantom{.} t& \quad\mbox{central spin}\\
\\
\dis \widetilde{t}=t &\quad \mbox{BCS}\\
\end{array}
\label{ttilde}
\en
No re-scaling is necessary for the BCS model. Further, analyzing equations of motion for $x$ and $y$ components of the total spin \re{h0ev} and \re{bcsev}, we conclude that they coincide for original and new spins, i.e. 
$$
\widetilde{J}_-=J_-
$$
To express original $n$ (pseudo)spins ${\bf s}_j(t)=2\langle \hat {\bf K}_j(t) \rangle$ in terms of $m$ new spins ${\bf \widetilde{s}}_j(\widetilde{t})$, we use equations \re{ratios} and \re{st} to rewrite the matrix element $B(u)$ given by equation \re{B} as  
\beg
\begin{array}{lcl}
\dis B(u)&=&\dis J_-\prod_{j=1}^{m-1}(u-u_j) \sum_{q=0}^{n-1} \frac{C_q}{u-\eps_q}=\\
\\
\dis &\phantom{=}&\dis \widetilde{B}(u)\widetilde{P}_m(u)\sum_{q=0}^{n-1} \frac{C_q}{u-\eps_q}\\
\end{array}
\label{Bratios}
\en
Evaluating residues at $u=\eps_j$ in equation \re{Bratios}, we obtain according to equation \re{res}  
\beg
s_j^-(t)=C_j \widetilde{P}_m(\eps_j) \sum_{q=0}^{m-1} \frac{ \widetilde{s}_j^-(\widetilde{t})}{\eps_j-\widetilde{\eps}_q}
\label{SS}
\en
where constants $C_j$ are given by equation \re{c}. Similarly, one can relate  $z$-components of original spins to new ones
\beg
s_j^z(t)=C_j \widetilde{P}_m(\eps_j)\left(-\alpha +\sum_{q=0}^{m-1}\frac{ \widetilde{s}_q^z(\widetilde{t})}{\eps_j-\widetilde{\eps}_q}\right)
\label{SSz}
\en
The solution for  new spins ${\bf \widetilde{s}}_j(\widetilde{t})$ as functions of time can be read off directly from the general solution -- equations \re{trans}, \re{Jt}, and \re{centr} of the previous section. We only need to replace $n$ with $m$, re-scale the time variable for the central spin model according to equation \re{ttilde}, and replace the spectral polynomial $Q_{2n}(u)$ with its reduced counterpart $\widetilde{Q}_{2m}(u)$ in the definition of matrices of periods \re{periods} of hyperelliptic functions.

For future references, let us also write down equations for components of original spins ${\bf s}_j(t)=2\langle \hat {\bf K}_j(t) \rangle$ and evolution of $J_-(t)$ for degenerate solutions in terms of $m$ variables $\widetilde{u}_j=u_j$.
Evaluating residues at $u=\eps_j$ in equation \re{Bratios} and using equation \re{res}, we find
\beg
s_j^-=C_jJ_-\prod_{j=1}^{m-1}(\eps_j-u_j)
\label{remain}
\en
With the help of equation \re{red1} and \re{remain}, evolution equations \re{h0ev} and \re{bcsev} for $J_-$ become
\beg
\begin{array}{l}
\dis \frac{d\ln J_-}{dt}=\frac{(-1)^m ie_0}{\sqrt{\widetilde {\alpha}_0}} \prod_{j=1}^{m-1}u_j\quad\mbox{central spin}\\
\\
\dis \frac{d\ln J_-}{dt}=i\biggl(\widetilde{\alpha}_{2m-1}+2\sum_{j=1}^{m-1} u_j\biggr)\quad\mbox{BCS}\\
\end{array}
\label{Jbcs}
\en
We  see that degenerate solutions with $m<n$  frequencies can be parameterized by $m$ auxiliary spins both for the BCS and central spin model. Therefore, corresponding trajectories live on an $m$-dimensional torus. By symmetry, for the same initial conditions, trajectories of all Hamiltonians $H_q$ in system \re{class} live on the same torus.  This implies, for example, that a slight change of initial values of integrals of motion \re{values} will induce small oscillations along the remaining $n-m$ directions. 

Let us consider several examples of degenerate solutions.\\
\noindent {\bf One spin solutions,} $\bf m=1$.  These solutions reproduce the BCS solution for the ground state and excitations of the BCS model (see also the discussion following equation~\re{bcs}).

Since the residual spectral polynomial is positively defined, we can parameterize it as     
$\widetilde{Q}_2(u)=(u-\mu)^2+\Delta_0^2$. For $m=1$, there is only one condition in equation \re{gap1} 
\beg
\sum_{q=0}^{n-1}\frac{e_q}{ \sqrt{(\eps_q-\mu)^2+\Delta_0^2 } }=\alpha \quad e_q=\pm1
\label{gap2}
\en
For the BCS model $\alpha=2/g$ and we recognize equation \re{gap2} as  the BCS gap equation. Similarly, equation \re{jz} becomes  the BCS equation for the chemical potential. 

The energy of  BCS model \re{ebcs} for $m=1$ is
\beg
{\cal E}_{BCS}=\frac{\Delta_0^2}{g}-\mu J_z-\sum_{j=0}^{n-1}e_j \sqrt{(\eps_j-\mu)^2+\Delta_0^2}
\label{gaudons}
\en
We see that a choice of signs $e_j=1$ for all $j$ yields the BCS ground state energy, while if we choose one of  $e_j$ to be negative, $e_q=-1$ and $e_j=1$ for $j\ne q$, the solution corresponds to a pair excitation of energy $2\sqrt{(\eps_q-\mu)^2+\Delta_0^2}$. Similarly, one obtains solutions with two and more excitations by choosing several signs $e_j$ to be negative.

According to equations \re{htilde}, the BCS and central spin models for $m=1$ reduce to the following single spin models:
\beg
\widetilde{H}_0=-\B\widetilde{s}_0^z\qquad \widetilde{H}_{BCS}=-\frac{g}{2} \widetilde{s}_0^+\widetilde{s}_0^-
\label{hnew}
\en
Then, equation \re{tildev} provides an alternative parametrization of the residual spectral polynomial $\widetilde{Q}_2(u)$
\beg
\widetilde{Q}_2(u)=u^2-\frac{2\widetilde{s}_0^z}{\alpha} u+ \frac{ {\bf \widetilde{s}}_0^2}{\alpha^2}
\label{Qspins}
\en
The solution of equations of motion is
\beg
J_-(t)=\widetilde{s}_0^-(\tilde t)=\alpha\Delta_0 e^{-i\beta t+i\phi}
\label{delta}
\en
where the frequency $\beta$ is different for BCS ($\beta_{BCS}$) and  central spin ($\beta_0$) models
$$
\beta_{BCS}=2\mu\quad \beta_0=\frac{-e_0}{\sqrt{\mu^2+\Delta_0^2}}\quad e_0=\pm 1
$$
Components of original spins ${\bf s}_j(t)=2\langle \hat {\bf K}_j(t) \rangle$ are given by equations \re{SS} and \re{SSz}. We have
\beg
s_j^-(t)=\frac{e_j J_-(t)}{\alpha\sqrt{(\eps_j-\mu)^2+\Delta_0^2}}\quad s_j^z=-\frac{e_j (\eps_j-\mu)}{\sqrt{(\eps_j-\mu)^2+\Delta_0^2}}
\label{bcsspins}
\en
Note that one spin solutions constructed above do not minimize  central spin Hamiltonian \re{csclass}. Instead, we observe that in this case the configuration of spins with minimum energy is $s_j^z=-\mbox{sgn}\gamma_j$. Similarly, we see from equations \re{class} that configurations that minimize Hamiltonians $H_q$ are  $s_j^z=-\mbox{sgn}(\eps_q-\eps_j)$. Interestingly, according to the definition of pseudospins \re{pseudo}, one of these configurations is the unperturbed Fermi ground state. Since all spins in these configurations are oriented along the $z$-axis, they are also stationary with respect to  BCS Hamiltonian \re{bcs}.

\noindent {\bf Two spin solutions,} $\bf m=2$. In this case there are two constraints \re{gap1} on  four coefficients of a positively defined polynomial $\widetilde{Q}_4(u)$. 
$$
\begin{array}{l}
\dis \sum_{q=0}^{n-1}\frac{e_q}{ \sqrt{\widetilde{Q}_{4}(\eps_q)} }=0\quad \sum_{q=0}^{n-1}\frac{e_q\eps_q}{ \sqrt{\widetilde{Q}_{4}(\eps_q)} }=\alpha\quad e_q=\pm 1\\
\\
\dis \widetilde{Q}_{4}(u) =\sum_{k=0}^{4} \widetilde{\alpha}_k u^k \quad \widetilde{\alpha}_{4}=1\\
\end{array}
$$
Equations \re{E0} and \re{ebcs} show that the energy of these solutions can take arbitrary values in the range of energies allowed for  the BCS and central spin models. 

The equivalent two spin problems \re{htilde} are
\beg
\widetilde{H}_{BCS}= 2\widetilde{\eps}_1 \widetilde{s}_1^z-\frac{g}{2} \widetilde{J}_+\widetilde{J}_- \quad \widetilde{H}_0=\frac{\widetilde{\gamma}_0}{2} \widetilde{{\bf s}}_0\cdot{\widetilde{\bf s}}_1-\B \widetilde{s}_0^z
\label{2spin}
\en
We are left with only one separation variable $u_1=\widetilde{u}_1$ and equations \re{jip} now read
\beg
\begin{array}{l}
\dis \dot u_1^2 +4\widetilde{Q}_4(u_1)=0 \quad\mbox{BCS}\\
\\
\dis \widetilde{\alpha}_0\dot u_1^2+\widetilde{Q}_4(u_1)=0\quad \mbox{central spin}\\
\end{array}
\label{u1}
\en
while for the total (pseudo)spin, using equation \re{Jbcs}, we obtain
\beg
\begin{array}{l}
\dis \frac{d\ln J_-}{dt}=\pm\frac{i}{\sqrt{\widetilde {\alpha}_0}}u_1\quad\mbox{central spin}\\
\\
\dis \frac{d\ln J_-}{dt}=i\widetilde{\alpha}_3+2iu_1\quad\mbox{BCS}\\
\end{array}
\label{J2spin}
\en
  Components of individual spins can be expressed either through $u_1$ and $J_-$  or in terms of new spins using equations \re{remain} and \re{SS} 
\beg
\dis s_j^-=C_jJ_-(\eps_j-u_1)=C_j(\eps_j-\widetilde{\eps}_1)\widetilde{s}_0^-(\widetilde{t})+C_j\eps_j \widetilde{s}_1^-(\widetilde{t})
\label{comp}
\en
A special case of two spin solutions when the four coefficients of a positively defined polynomial $\widetilde{Q}_4(u)$ can be parameterized
by three numbers, $\widetilde{Q}_4(u)=((u-\omega)^2+\Delta_-^2+\Delta_+^2)^2-4\Delta_-^2\Delta_+^2$,  has been previously discovered in Ref.~\onlinecite{barankov}. The general two spin solution is obtained from equations (\ref{u1}-\ref{comp}), which are solved by setting $n=2$ ($\g=1$)  in equations \re{R}, \re{trans} and \re{Jt} where  hyperelliptic functions now become ordinary elliptic functions. 

Similarly, one can obtain  three, four etc. spin solutions (degenerate solutions with $m=3,4$ etc. frequencies) in terms of hyperelliptic functions of genus $\g=2, 3$ etc. 
However, as discussed in the beginning of this section, solutions with $m<n$ frequencies are nonrepresentative of the dynamics anywhere in the phase space and are expected to be unstable in the sense of KAM theorem. 

Moreover, we emphasize that degenerate solutions derived here are not all solutions with $m<n$ independent frequencies. Indeed, it is clear that for a generic point in the phase space all roots of the spectral polynomial $Q_{2n}(u)$ are distinct and,  by continuity, the distance to the closest point where two roots coincide is finite. On the other hand, there are points with any number of frequencies arbitrarily close to  any point in the phase space.  Other solutions with $m<n$ frequencies can  be constructed  using the reduction theory for hyperelliptic functions\cite{victor1}. 

Note also that degenerate solutions with $m$ frequencies contain ones with fewer frequencies and can be further reduced to  $m-1$, $m-2$ etc. Geometrically, they live on a $2m$-dimensional surface in the $2n$-dimensional phase space (see the remark below equation \re{Jbcs}). Surfaces with smaller $m$ are embedded into ones with larger $m$. Interestingly, periodic trajectories that correspond to the BCS energy spectrum ($m=1$) belong to all these surfaces.

\section{Applications and Open Problems}
\label{open}

Our results can be directly applied to the problem of decoherence within the central spin model. In this case, we believe, a complete exact answer can be obtained using asymptotics of hyperelliptic functions. 

Another possible application is to experiments\cite{tarucha} on electron transport in spin blockaded semiconductor double dots. In this connection, it might be useful to analyze first a simpler setup of a single dot connected to two polarized leads. This setup leads\cite{glazman} to an integrable model very similar to the central spin model.

Further, it might be possible to use the BCS  model \re{bcs} with few classical spins to describe a number of grains connected to each other by Josephson junctions.

Other interesting applications include pairing phenomena in cold fermion gases\cite{feshbach_experiments} modeled either by BCS or Dicke model and in superconducting metals (see e.g. the discussion in Ref.~\onlinecite{galperin}). In these cases, a careful identification of observable robust features of the solution is needed.

An interesting open problem is the evaluation of leading finite size (quantum) corrections to the general solution obtained in this paper.

\section{Acknowledgements}

We thank L. Glazman and L. Levitov for drawing our attention to  the problem of determining the dynamics of the central spin and BCS models and for pointing out a number of  relevant references. We are also grateful to them and to A. Lamacraft, A. Polyakov, N. Saulina, and P. Wiegmann for interesting discussions.


\begin{thebibliography}{99}
\bibitem{argum} If the trap has certain spatial symmetries, the coupling $g$ in \re{bcs1} can depend on $j$ and $q$.
Hamiltonians \re{bcs1}  and \re{bcsmomentum} neglect any such effects, which are usually not important (see also the discussion of stability against perturbations later in the paper).
\bibitem{field} Since the component of the total spin along the magnetic field is conserved, including a uniform magnetic field acting on enviromental spins is equivalent to adding a constant to the Hamiltonian.
\bibitem{product} Within the mean field approximation, this is always true in equilibrium with any value of the coupling constant.
\bibitem{pair} These are excitations that conserve the number of Cooper pairs, see Refs.~\onlinecite{BCS,ander1}.
\bibitem{lax} Matrix ${\cal L}$ is called the Lax matrix of integrable system \re{class}.
\bibitem{genus} The genus is the number of handles on the Riemann surface of $z(u)=\sqrt{Q_{2n}(u)}$. The latter consists of two branches glued along $n$ cuts (see Fig.~1). Each branch is a complex plane that is equivalent to a sphere. Because the two spheres are glued along $n$ cuts, there are $\g=n-1$ handles.
\end{thebibliography}
\end{document}